\documentclass[twocolumn,showpacs,showkeys,preprintnumbers,aps]{revtex4-1} 
\usepackage{amssymb}
\usepackage{amsfonts}
\usepackage{latexsym}
\usepackage{dcolumn}
\usepackage{amsmath}
\usepackage{graphicx}
\usepackage{psfrag,pstricks}
\usepackage{float}
\usepackage{bm}
\newcommand{\la}{\langle}
\newcommand{\ra}{\rangle}
\begin{document}
\title{Response of a mechanical oscillator in an optomechanical cavity driven by a finite-bandwidth squeezed vacuum excitation} 
\author{H. Lotfipour} \email{hodalotfipoor@gmail.com}

\affiliation{Institute of Laser and Plasma of Shahid Beheshti University, Tehran, Iran}
\author{S. Shahidani}

 \author{R. Roknizadeh} \email{r.roknizadeh@gmail.com} 
\ \author{M. H. Naderi}
 \affiliation{
 Department of Physics, Quantum Optics Group\\University of Isfahan, Isfahan, Iran}

\date{\today}

\begin{abstract}
In this paper, we theoretically investigate the displacement and momentum fluctuations
spectra of the movable mirror in a standard optomechanical system driven by a finite-bandwidth
squeezed vacuum light accompanying a coherent laser field. Two cases in which the squeezed vacuum is generated by degenerated and non-degenerate parametric oscillators (DPO and NDPO) are considered.
We find that for the case of finite-bandwidth squeezed vacuum injection, the two spectra
exhibit unique features, which strongly differ from those of broadband squeezing
excitation. In particular, the spectra exhibit a three-peaked and a four-peaked structure,
respectively, for the squeezing injection from DPO and NDPO. Besides, some anomalous
characteristics of the spectra such as squeezing-induced pimple, hole burning, and
dispersive profile are found to be highly sensitive to the squeezing parameters and the
temperature of the mirror.
We also evaluate the mean-square fluctuations in position and momentum quadratures of the movable mirror and analyze the influence of the squeezing parameters of the input field on the mechanical squeezing. It will be shown that the parameters of driven squeezed vacuum affects the squeezing. We find the optimal mechanical squeezing is achievable via finite-bandwidth squeezed vacuum injection which is affected by the intensity of squeezed vacuum. We also show that the phase of incident squeezed vacuum determines whether position or momentum squeezing occurs. 
Our proposed scheme not only provides a feasible experimental method to detect and characterize squeezed light by optomechanical systems, but also suggests a way for controllable transfer of squeezing from an optical field to 
a mechanical oscillator.
\end{abstract}

\pacs{42.50.Wk, 42.50.Lc, 42.65Lm }
\keywords {finite-bandwidth squeezed vacuum field, cavity optomechanics, quantum fluctuations, mechanical squeezing}
\maketitle
\section{introduction}\label{sec1}
Over the past decade, we have witnessed enormous and rapid progress in the field of
cavity quantum optomechanics, the field of research exploring the coupling of optical
radiation to mechanical motion from both theoretical and experimental points of view (for
a recent review, see, e.g., \cite{Aspelmeyer}). This field has emerged as an ideal platform to explore the
applicability of quantum mechanics to systems of much larger sizes and masses than the
atomic and particle scales thanks to sophisticated experiments , including the cooling of
the mechanical motion down to the quantum ground state \cite{Teufel,Chan,Verhagen}, the detection of quantized mechanical motion \cite{Safavi1,Brahms}, coherent state transfer between cavity and mechanical modes \cite{Zhou,Palomaki}, the realization of squeezed light \cite{Brooks,Safavi2}, and the preparation of mechanical squeezed state \cite{Wollman,Pirkkalainen,Lecocq}. Interestingly, the exploration of quantum features in optomechanical systems has not only led to the development of novel applications, but also opened new insights into the fundamental properties of nature. The examples include precision measurements \cite{Verlot,Zhang,Wang}, the development of hybrid systems \cite{Rabl1}, probing open quantum system dynamics \cite{Vanner}, quantum information processing \cite{Stannigel}, and probing the interface between quantum mechanics and gravity \cite{Pikovski}.\\

  Quantum squeezing of a mechanical oscillator, characterized by an uncertainty of a single
motional quadrature (position or momentum) beyond the standard quantum limit, is one
of the key macroscopic quantum effects that can be utilized to investigate the quantum to classical
transition and to improve the precision of quantum measurements such as the
detection of gravitational waves \cite{Caves,Abramovici}. Although only a few experimental
realizations have been reported very recently \cite{Szorkovszky1,Pontin,Vinante,Wollman,Pirkkalainen,Lecocq}, many proposals have been put forward to generate and enhance mechanical
squeezing in optomechanical systems. Some examples include the conditional quantum
measurements \cite{Clerk,Ruskov}, parametric amplification \cite{Mari,Huang1,Szorkovszky,Liao,Kronwald,Asjad,Lu,Benito}, coupling a nanomechanical
oscillator to an atomic gas \cite{Ian}, quantum reservoir engineering \cite{Gu1,Tan,Rabl}, exploiting the
periodically-modulated driving on the dissipative optomechanical system \cite{Gu2,Vanner}, and squeezing via intracavity nonlinear crystal \cite{Otey,Agarwal}.\\ 

Research on the interaction of squeezed light with matter has been one of the most attractive issues in quantum optics over the past many years. In particular, considerable attention has been directed at modifying the radiative properties of atom via interaction with a squeezed light. The basis for this attention originates from the prediction of a sub-natural linewidth in the spontaneous emission spectrum of a two-level atom in a broadband squeezed vacuum bath  \cite{Gardiner1}. This effect results from the quantum correlations between pairs of photons in the squeezed vacuum, produced, e.g., by the process of parametric down-conversion, which lead to reduced quantum fluctuations in one quadrature of the field driving the atom. Following from this prediction, a number of other interesting and novel quantum effects, arising from quantum correlations and noise reduction in the squeezed vacuum interacting with an atom, have been studied. Some of these include sub-natural linewidth in resonance fluorescence spectrum \cite{Carmichael} as well as in weak field absorption spectrum \cite{Ritsch}, population trapping \cite{Cabrillo}, anomalous resonance fluorescence \cite{Smart,Swain0,Swain1}, hole burning and dispersive profiles in the probe absorption spectrum \cite{ Zhou1}, and linear two-photon excitation \cite{Ficek}. The above-cited studies have been carried out assuming the squeezed vacuum to be broadband, i.e., its width to be much larger than the atomic linewidth and the Rabi frequency of the driving field. However, experimental realizations of squeezed light by subthreshold optical parametric oscillators \cite{Wu,Georgiades,Polzik} indicate that the bandwidth of the squeezed light is typically of the order of the atomic linewidth. In this sense, some other studies have been performed to explore the response of a  two-level atom to a squeezed vacuum excitation with finite-bandwidth\cite{Gardiner2,Parkins,Ritsch,Tanas,Messikh,Tesfa}. The results reveal that the atomic dynamics,  radiative properties, and photon statistics of the emitted radiation exhibit unique features which do not appear for a broadband excitation.\\

       In comparison with atomic systems, the interaction of mechanical oscillators with squeezed light has not been investigated extensively. The relevant investigations are mainly focused on the generation and enhancement of mechanical squeezing in an optomechanical cavity by injecting finite-bandwidth \cite{Jahne} or broadband\cite{Huang2,Gu1} squeezed vacuum light into the cavity. Aside from the generation of mechanical squeezing, injecting the optomechanical systems with squeezed light may lead to the entanglement between two separate nanomechanical oscillators \cite{Huang3,Sete} and electromagnetically induced transparency \cite{Huang4}. It should be noted that in atomic systems, quantum coherent control of mechanical motion is state of the art \cite{Blatt}. In contrast, the fabricated nanomechanical and micromechanical resonators extend this level of control to a different realm, of objects with large masses and of devices with a great flexibility in design and the possibility to integrate them in on-chip architectures.\\
      In this paper, the response of a movable mirror in an optomechanical cavity to DPO and NDPO finite-bandwidth squeezed vacuum states as input probe fields is investigated and the results are compared to the case of broad-band squeezed vacuum injection. In particular, the effects of the bandwidth and squeezing parameters of the squeezed vacuum input on the displacement and momentum fluctuations spectra as well as the mechanical squeezing of the movable mirror are analyzed. We show that a squeezed vacuum of bandwidth smaller than the cavity decay rate induces certain effects that are unique to finite-bandwidth excitations and the quantum nature of squeezed light.
We show some anomalous features such as pimple, hole burning, and dispersion-like profile in the spectra of the movable mirror which has not yet been studied in optomechanical systems.\\

  We also study the role of mirror temperature in the appearance or suppression of these features. We find that when  hole burning appears, the two-photon correlation of the incident squeezed vacuum is transferred to the spectral density of movable mirror, and accordingly one can use the optomechanical cavity for detecting the two-photon correlation in the driving squeezed vacuum.\\ 

 We also examine the squeezing of the position and momentum quadratures of the movable mirror and analyze how the mechanical squeezing is affected by the squeezing parameters as well as the type (DPO or NDPO) of the squeezed vacuum input. The results reveal that the maximum mechanical squeezing occurs for the case of finite-bandwidth DPO squeezed vacuum input.\\
     The remainder of the paper is structured as follows. In Sec. \ref{sec2}, we introduce the physical model of the system under consideration, give the quantum Langevin equations, and obtain the steady-state mean values of the relevant dynamical variables. In Sec. \ref{sec3}, we consider the spectra of small fluctuations in the position and momentum quadratures of the oscillating mirror and then, in Sec. \ref{sec4}, we derive the analytical forms of the mirror displacement and momentum spectra for both cases of DPO and NDPO squeezed vacuum input fields. We devote Sec. \ref{sec5} to analyze in detail the displacement and momentum spectra as well as the mechanical squeezing of the movable mirror. Finally, we summarize our conclusions in Sec. \ref{sec6}. 
  
 \section{The physical  model}\label{sec2}
\begin{figure}[ht]
\includegraphics[width=\columnwidth]{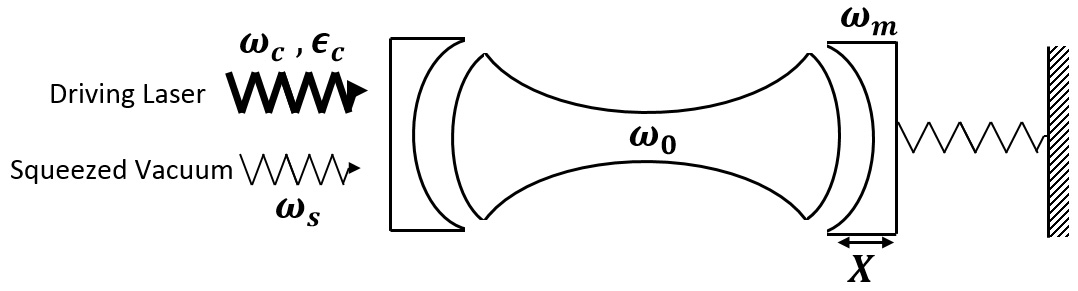}
\vspace*{-5mm}
\caption{\footnotesize Schematic description of the system under consideration. An oscillating mirror with frequency $ \omega_{m} $ is coupled via radiation pressure to the cavity field of frequency $ \omega_{0} $. The cavity is driven by a laser of frequency $ \omega_{c} $, accompanying a much weaker squeezed vacuum field with central frequency $ \omega_{s} $. }\label{schematic}
\end{figure}

As depicted in Fig.\ref{schematic}, we consider a standard optomechanical cavity where the cavity mode is an optical harmonic oscillator with frequency $ \omega_{0} $,
   coupled to an oscillating mirror with frequency $ \omega_{m} $ and damping rate $ \gamma_{m} $.
 The cavity mode  is driven by a strong pump laser field of frequency $ \omega_{c} $ and  amplitude $ \varepsilon _{c}$ through the fixed mirror. We further assume that the cavity is fed with a weak squeezed vacuum field at frequency $ \omega_{s}=\omega_{c}+\omega_{m} $. The Hamiltonian of the system in a reference frame rotating at the laser frequency can be written as
\begin{equation}
H=\hbar\Delta _{0}a^{\dagger}a+\dfrac{\hbar\omega _{m}}{2}(p^{2}+q^{2})
-\hbar g_{0}a^{\dagger}aq+i\hbar\varepsilon _{c}(a^{\dagger}-a).\label{1}
\end{equation}
In the above Hamiltonian, the first and the second terms, respectively, indicate the cavity mode energy (described by the creation and annihilation operators $ a^{\dagger} $ and $ a $) and the mechanical mode energy (described by the  dimensionless displacement and momentum operators $ q $ and $ p $ and mass $m$). The third term describes the  interaction between the  mechanical oscillator and the cavity mode with single-photon coupling strength  $ g_{0}=\dfrac{\omega _{0}}{L}\sqrt{\dfrac{\hbar}{2m\omega _{m}}} $. Here, $ L $ is the cavity length in mechanical equilibrium.  Finally, the fourth term corresponds to  the driving of the intracavity mode with the input laser. We also introduce the amplitude of the pump field,  $ \varepsilon _{c}=\sqrt{\dfrac{2\kappa P}{\hbar\omega _{c}}} $ where $ \kappa $ is the cavity decay rate through its input port and $ P $ is the input laser power. Moreover, $ \Delta_{0}=\omega _{0}-\omega _{c} $denotes the  cavity-pump detuning.

The full dynamics of the system is described by the set of the following nonlinear quantum Langevin equations 
\begin{subequations}\label{langevin}
\begin{eqnarray}
\dot{q}&=&\omega _{m}p\label{2a},\\ 
\dot{p}&=&-\omega _{m}q+g_{0}a^{\dagger}a-\gamma _{m}p+\xi \label{2b},\\ 
\dot{a}&=&-i\Delta _{0}a+ig_{0}qa+\varepsilon _{c}-\kappa a+\sqrt{2\kappa}a_{in}\label{2c},
\end{eqnarray}
\end{subequations}
where $ \xi $ is the Brownian noise operator which describes the heating of the mirror by its thermal environment at temperature $T$ and 
 $ a_{in} $  is  the optical input noise operator.

The steady-state solutions of the quantum Langevin equations in the classical limit can easily be found to be
\begin{subequations}\label{steady}
\begin{eqnarray}
a_{s}&=&\dfrac{\kappa -i\Delta}{\Delta ^{2}+\kappa ^{2}}\varepsilon _{c},\\ 
q_{s}&=&\dfrac{g_{0}}{\omega _{m}}\vert a_{s}\vert ^{2},\\ 
p_{s}&=&0,
\end{eqnarray}
\end{subequations} 
where $ \Delta =\Delta _{0}-g_{0}q_{s} $ is defined as the effective detuning of the cavity.
 \section{Small fluctuation dynamics}\label{sec3}
Since we consider the pump laser field is strong and the squeezed vacuum field is weak enough, the classical solution is a strong value and the quantum mechanics acts like a damped fluctuations (noise), therefore, we can use the linearized description to examine the  fluctuation  dynamics of the oscillating mirror under the influence of the input noises and decompose each operator in Eqs.(\ref{langevin}a)-(\ref{langevin}c)
as the sum of its classical steady-state value ,given by the set of Eqs. (\ref{steady}a)-(\ref{steady}c), and a small fluctuation,
\begin{equation}\label{linearized}
a=a_{s}+\delta a,\quad q=q_{s}+\delta q,\quad
p=p_{s}+\delta p.
\end{equation}
In this manner the linearized quantum Langevin equations for the fluctuation operators can be written in the compact matrix form
\begin{equation}\label{matrix}
\dot{u}=Mu(t)+n(t),
\end{equation}
where the vector of fluctuation operators is $u(t)=(\delta q, \delta p ,\delta x , \delta y)^{T} $, and the corresponding vector of noises is  given by $ n(t)= (0 , \xi , \sqrt{2\kappa} \delta x_{in} , \sqrt{2\kappa}\delta y_{in})^{T}  $. Here, we have defined 
the  cavity-field quadratures as $ \delta x=(\delta a+\delta a^{\dagger})/\sqrt{2} $ and $ \delta y=i(\delta a^{\dagger}-\delta a)/\sqrt{2} $ and the input noise quadratures as $ \delta x_{in}=(\delta a_{in}+\delta a^{\dagger}_{in})/\sqrt{2} $ and $\delta y_{in}=i(\delta a^{\dagger}_{in}-\delta a_{in})/\sqrt{2} $.  Furthermore,  the drift matrix $ M $ is given by
\begin{equation}\label{M}
M=\left(
\begin{array}{cccc}
0 & \omega _m & 0 & 0\\ 
-\omega _m & -\gamma _{m} & g& 0 \\ 
0 & 0 &  -\kappa & \Delta  \\ 
g & 0 & -\Delta  &  -\kappa 
\end{array} \right),
\end{equation}
where  $ g=\sqrt{2}g_{0}a_{s} $ is the  light-enhanced optomechanical coupling for the linearized regime. 
The steady state associated with Eq. (\ref{matrix}) is reached when the
system is stable, which occurs if and only if all the eigenvalues
of the matrix $ M $ have a negative real part. These stability conditions can be obtained, for example, by using the Routh-Hurwitz criteria \cite{Hurwitz}.
Since we are interested in the spectrum of fluctuations in displacement and momentum of the movable mirror, it is more convenient to work in the frequency domain. To this end, we write Eq. (\ref{matrix}) in the Fourier space by using 
\begin{equation}\label{f}
f(t)=\dfrac{1}{2\pi}\int^{+\infty}_{-\infty} d\omega e^{-i\omega t}f(\omega),
\end{equation}
\begin{equation}\label{fdagger}
f^{\dagger}(t)=\dfrac{1}{2\pi}\int^{+\infty}_{-\infty} d\omega e^{-i\omega t}f^{\dagger}(-\omega),
\end{equation}
and solve it to get the following expressions for the displacement and momentum fluctuations of the movable mirror

\begin{subequations}\label{deltaq}
\begin{eqnarray}
&\delta q(\omega)=&\nonumber \\&F_{1}(\omega)&\xi(\omega)+F_{2}(\omega)\delta a^{\dagger}_{in}(-\omega)+F_{3}(\omega)\delta a_{in}(\omega),\\
&\delta p(\omega)=&\nonumber \\&E_{1}(\omega)&\xi(\omega)+E_{2}(\omega)\delta a^{\dagger}_{in}(-\omega)+E_{3}(\omega)\delta a_{in}(\omega),
\end{eqnarray}
\end{subequations}

where
\begin{subequations}\label{F ha}
\begin{eqnarray}
F_{1}(\omega)&=&\dfrac{\omega _{m}}{d(\omega)}\lbrace(\kappa -i\omega)^{2}+\Delta^{2}\rbrace,\\ 
F_{2}(\omega)&=&\dfrac{g\omega _{m}\sqrt{\kappa}}{d(\omega)}\lbrace{\kappa +i(\Delta -\omega)}\rbrace,\\ 
F_{3}(\omega)&=&F^{*}_{2}(-\omega),\\
E_{l}(\omega)&=&-\dfrac{i\omega}{\omega_{m}}F_{l}(\omega)\quad (l=1,2,3),\\
d(\omega)&=&[\Delta ^{2}+(\kappa -i\omega)^{2}]\nonumber\\&\times &(\omega ^{2}_{m}-\omega ^{2}-i\omega\gamma _{m})-g^{2}\omega _{m}\Delta.
\end{eqnarray}
\end{subequations}
In each of Eqs. (\ref{deltaq}a) and (\ref{deltaq}b), the first term involving $ \xi(\omega) $ originates from the thermal noise of the movable mirror, while the other two terms involving the contribution of the optical
input noise $ \delta a_{in}(\omega) $ arise from the radiation pressure. In the absence of the radiation
pressure coupling, the moving mirror undergoes pure Brownian motion with a Lorentzian
shape susceptibility with width $ \gamma_{m}$ centered about $ \omega_{m} $. The optical input noises cause
changes in both the width and central frequency of the susceptibility and imprint
themselves on the displacement and momentum spectra of the moving mirror.\\

The symmetrized spectrum of the displacement and momentum fluctuations of the
movable mirror is given by \cite{Huang5}
\begin{align}\label{SF}
S_{F}(\omega)=\dfrac{1}{2}[S_{FF}(\omega)+S_{FF}(-\omega)],\quad F=p,q
\end{align}
where $ S_{FF}(\omega) $ is the Fourier transform of the two time correlation functions $ \la \delta F(t)\delta F(0)\ra  $

\begin{align}\label{SOmega}
S_{FF}(\omega)=\int _{-\infty}^{+\infty}dte^{i\omega t}\la \delta F(t)\delta F(0)\ra , \quad F=p,q.
\end{align}

 To determine $ S_{F}(\omega) $, we require the correlation functions of the noise sources in the frequency domain which will be calculated in the next section.
\section{Optomechanical system driven by squeezed vacuum excitation}\label{sec4}
To study the response of the optomechanical system to the driving squeezed vacuum field, it is required to calculate  two physical outputs: the spectral density and the mean square of fluctuations of the displacement and momentum of the movable mirror.\\

In the system under investigation, the squeezed vacuum source is assumed to be either
a DPO or a NDPO. The output fields from DPO and NDPO are characterized by the following
correlation functions \cite{Collet}: 

\begin{subequations}\label{squeezed correlation} 
\begin{flalign}
&\langle  \delta a_{out}(\omega)\delta a_{out}(\Omega)\rangle =2\pi M(\omega) \delta(2\omega _{s}-\omega -\Omega ),\\
&\langle  \delta a^{\dagger}_{out}(-\omega)\delta a^{\dagger}_{out}(-\Omega)\rangle =  2\pi M^{*}(-\omega)\delta(2\omega _{s}+\omega +\Omega ),\\ 
&\langle \delta a^{\dagger}_{out}(-\omega)\delta a_{out}(\Omega)\rangle = 2\pi N(-\omega) \delta(\omega +\Omega ),\\
&\langle \delta a_{out}(\omega)\delta a^{\dagger}_{out}(-\Omega)\rangle = 2\pi (N(\omega)+1) \delta(\omega +\Omega ),
\end{flalign}
\end{subequations} 
where $ N(\omega) $ is related to the mean number of photons at frequency $  \omega $, while
$ M(\omega) $ is characteristic of the squeezed vacuum field and describes the correlation between
the two photons created in the down-conversion process. Furthermore, the
frequencies $  \omega $ and $ \Omega $ are measured with respect to a certain given central frequency. The
photon number and two-photon correlation functions are not independent of each other but can
be shown to satisfy the inequality $ \vert M(\omega)\vert\leq\sqrt{N(\omega)[N(\omega)+1]} $ . In the case of a coherent (ideal)
squeezed state, such as that produced by an optical parametric oscillator, the equality
holds. The frequency dependence of $ N(\omega) $ and $ M(\omega) $  in the output of an optical parametric oscillator, below of the threshold, for the ideal DPO is given by \cite{Collet} 

\begin{equation}
N(\omega)=\dfrac{\lambda ^{2}-\mu ^{2}}{4} [\dfrac{1}{(\omega -\omega _{s})^{2}+\mu ^{2}}-\dfrac{1}{(\omega -\omega _{s})^{2}+\lambda ^{2}}],\label{N}
\end{equation}
\begin{equation}
M(\omega)=e^{i\phi_{0}}\dfrac{\lambda ^{2}-\mu ^{2}}{4}[\dfrac{1}{(\omega -\omega _{s})^{2}+\mu ^{2}}+\dfrac{1}{(\omega -\omega _{s})^{2}+\lambda ^{2}}],\label{M}
\end{equation}
while for the NDPO we have \cite{Drummond} 
\begin{align}
&N(\omega)=\dfrac{\lambda ^{2}-\mu ^{2}}{8} (\dfrac{1}{(\omega -\omega _{s}-\alpha)^{2}+\mu ^{2}}+ \dfrac{1}{(\omega -\omega _{s}+\alpha)^{2}+\mu ^{2}} \nonumber\\ &- \dfrac{1}{(\omega -\omega _{s}-\alpha)^{2}+\lambda ^{2}}- \dfrac{1}{(\omega- \omega _{s}+\alpha)^{2}+\lambda ^{2}})\label{N1}
\end{align}
\begin{align}
&M(\omega)=e^{i\phi_{0}}\dfrac{\lambda ^{2}-\mu ^{2}}{8} \times \nonumber \\ &[\dfrac{1}{(\omega -\omega _{s}-\alpha)^{2}+\mu ^{2}}+\dfrac{1}{(\omega -\omega _{s}+\alpha)^{2}+\mu ^{2}}\nonumber \\ &+ \dfrac{1}{(\omega -\omega _{s}-\alpha)^{2}+\lambda ^{2}}+\dfrac{1}{(\omega -\omega _{s}+\alpha)^{2}+\lambda ^{2}}].\label{M1}
\end{align}
The parameters $ \lambda $ and $ \mu $ are related to the damping rate of the parametric oscillator cavity, $ \kappa _{p} $,
and the effective pump amplitude $ \epsilon $ of the coherent field driving the parametric oscillator
\begin{align}\label{lan}
\lambda =\dfrac{\kappa _{p}}{2}+\epsilon ,\quad\quad\quad \mu =\dfrac{\kappa _{p}}{2}-\epsilon.
\end{align}
and $ \phi_{0} $ is the phase of the pump field and \cite{Wu1}
\begin{align}\label{epsi}
\epsilon =\dfrac{E}{E_{c}}\dfrac{\kappa_{p}}{2},
\end{align}
where $ E $ is the amplitude of the pump coherent field and $ E_{c} $ is its threshold value for parametric oscillator. In OPO (Optical Parametric Oscillator), $ E $ is related to the power of pumping ($ P $) \cite{Drummond1}, so the effective pump amplitude is related to the main physical ratio of input pump power to the critical power, $ r=P/P_{c} $, and we have  $ \epsilon=\sqrt{r}\kappa_{p}/2 $.
The observed critical powers are in the range 10-15 mW (25-30 mW) for non-degenerate (degenerate) oscillation \cite{Wu1}. The noise spectrum and the squeezing level of the output light from OPO is related to $ \dfrac{\epsilon}{\kappa_{p}/2} $. When this ratio goes to 1 and therefore $ r\rightarrow1 $, the  threshold happens in OPO.

It is considerable that Eqs. (\ref{N})-(\ref{lan}) are found by applying standard linearization to the OPO, and they are only valid sufficiently below of the threshold, that is, when $0< \epsilon<\dfrac{\kappa_{p}}{2} $, both $ \lambda $ and $ \mu $ are positive and 
$ \lambda>\mu $, and the squeezing values are not too large.

 When the
parameters $ \lambda $ and $ \mu $ are much greater than all other relaxation rates in the problem, the
frequency dependence of $ N(\omega) $ and $ M(\omega) $ can be neglected. This case is referred to as
broadband squeezed vacuum in which there is no difference between the output fields
from DPO and NDPO.
The parameter  $ \alpha=\dfrac{(\omega _{1}-\omega _{2})}{2} $ represents the displacement from the central frequency of the
squeezing at which the two-mode squeezed vacuum is maximally squeezed. 

 The Brownian noise operator $ \xi $ associated with the coupling of the movable mirror to its thermal environment obeys the following
correlation function\cite{Huang5}
\begin{equation}
\langle \xi(\omega) \xi(\omega^{\prime})\rangle = 4\pi \dfrac{\gamma_{m}}{\omega_{m}}\omega[\textsl{coth} (\dfrac{\hbar\omega}{2k_{B}T})+1]\delta(\omega+\omega^{\prime}).
\label{thermal correlation}
\end{equation}

We are now in a position to determine the fluctuation spectra of the displacement and
momentum of the moving mirror. Considering the parametric oscillator output field,
characterized by the correlation functions (\ref{squeezed correlation}a-d), as the optical input noise to the
optomechanical cavity and combining Eqs. (\ref{SF})-(\ref{thermal correlation}), we arrive at the results

\begin{align}
S_{q}(\omega)=&\vert F_{1}(\omega)\vert^{2}\gamma_{m}\dfrac{\omega}{\omega_{m}}\textsl{coth}(\dfrac{\hbar\omega}{2k_{B}T})\nonumber\\ +&\vert F_{2}(\omega)\vert^{2}N(-\omega)+\vert F_{2}(-\omega)\vert^{2}N(\omega)\nonumber\\ +&\textbf{Re}[M^{*}(-\omega)F_{2}(\omega)F_{2}(-2\omega_{s}-\omega)\nonumber\\
+&M(\omega)F_{3}(\omega)F_{3}(2\omega_{s}-\omega)]\nonumber\\+
&\dfrac{1}{2}(\vert F_{2}(\omega)\vert^{2}+\vert F_{2}(-\omega)\vert^{2}),
\label{final S_{q}}
\end{align}
\vspace*{-5mm}
\begin{align}
S_{p}(\omega)=&(\dfrac{\omega}{\omega_{m}})^2\vert F_{1}(\omega)\vert^{2}\gamma_{m}\dfrac{\omega}{\omega_{m}}\textsl{coth}(\dfrac{\hbar\omega}{2k_{B}T})\nonumber\\ +&(\dfrac{\omega}{\omega_{m}})^2\dfrac{1}{2}(\vert F_{2}(\omega)\vert^{2}N(-\omega)+\vert F_{2}(-\omega\vert^{2}N(\omega))\nonumber\\ +&\textbf{Re}[\dfrac{-\omega(-2\omega_{s}-\omega)}{\omega_{m}^2}M^{*}(-\omega)F_{2}(\omega)F_{2}(-2\omega_{s}-\omega)\nonumber\\
+&\dfrac{\omega(\omega-2\omega_{s})}{\omega_{m}^2}M(\omega)F_{3}(\omega)F_{3}(2\omega_{s}-\omega)]\nonumber\\+
&\dfrac{1}{2}(\dfrac{\omega}{\omega_{m}})^2(\vert F_{2}(\omega)\vert^{2}+\vert F_{2}(-\omega)\vert^{2}).
\label{final S_{p}}
\end{align}
In each of the above two equations, the first term results from the thermal noise of the
movable mirror, the next two terms involving $ N(\omega) $ and $ M(\omega) $ originate from the squeezed
vacuum, and the last term is the contribution of the spontaneous emission of the input vacuum noise.

In order to investigate the quadrature squeezing of the moving mirror, we need to
evaluate the variances of its displacement and momentum operators. The mean square of
fluctuations of the displacement and momentum of the mirror are, respectively, calculated
as
\begin{subequations}\label{mean square}
\begin{eqnarray}
\langle \delta q^2(t)\rangle &=&\int_{-\infty}^{+\infty}\dfrac{d\omega}{2\pi} S_{q}(\omega)\label{21a},\\
 \langle \delta p^2(t)\rangle &=&\int_{-\infty}^{+\infty}\dfrac{d\omega}{2\pi} S_{p}(\omega) \label{21b}.
\end{eqnarray}
\end{subequations} 

The two quadratures $ q $ and $ p $ satisfy the commutation relation $  [q,p]=i$, which yields the
uncertainty relation $ \langle \delta q^2\rangle \langle \delta p^2\rangle \geq 1/4 $. The mirror motion is squeezed if either $  \langle \delta q^2\rangle $ or
$  \langle \delta p^2\rangle $ is less than $ 1/2 $ .

\section{Results and Discussions}\label{sec5}
\subsection{Displacement and momentum spectra of the moving mirror}

\begin{figure}[ht]
\includegraphics[width=150mm]{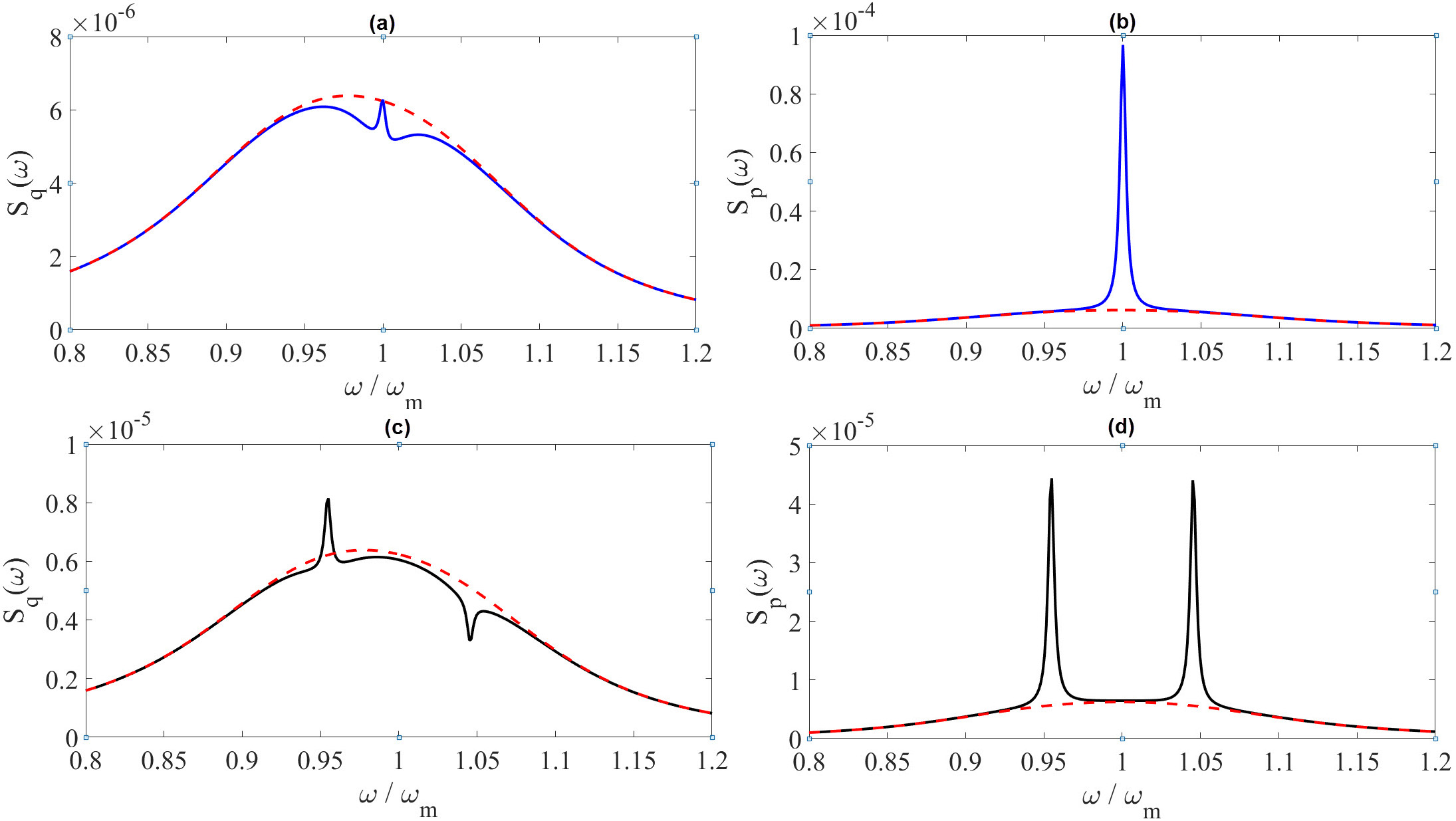}
\vspace*{-5mm}
\caption{\footnotesize (Color online) (a) The displacement and (b) the momentum spectrum (W/Hz) of the moving mirror for the case of DPO (blue line) and (c) the displacement and (d) the momentum spectrum (W/Hz) of the moving mirror
   for the case of NDPO with $ \alpha=2\kappa_{p} $ (black line) versus the normalized frequency $ \omega / \omega_{m} $ when $ T = 100 $ mK and $ P=5 $ mW. The spectrum(W/Hz) for the case of normal
vacuum input is plotted for comparison (red dashed line). The parameters of the squeezed vacuum input are $ \kappa_{p}=0.1\kappa,\epsilon=0.4\kappa_{p}$ and $\phi_{0}=0 $.}\label{plot1}
\vspace*{-1mm}
\end{figure}

In this section, by using Eqs. (\ref{F ha}a-e), (\ref{final S_{q}}), (\ref{final S_{p}}), and (\ref{N})-(\ref{M1}), we numerically evaluate and
analyze the spectra $ S_{q}(\omega) $ and $ S_{p}(\omega) $ to explore the effects of various physical parameters,
such as the input laser power $ P $, temperature $ T $, as well as the parameters associated with
the input squeezed vacuum field, i.e., $ \kappa_{p} $ , $ \phi_{0} $, and $ \alpha $ on the mechanical response of the
moving mirror. As we shall find in the following, by adjusting these parameters one can
effectively control the displacement and momentum fluctuation spectra. We analyze our
results based on the experimentally feasible parameters given in \cite{Groblacher}. We have,
in particular, $L=25$mm, $m=145$ ng, $ \kappa=2\pi\times215$kHz, $ \omega_{m}=2\pi\times947$kHz, and $ \gamma_{m}=2\pi\times141 $Hz.
Moreover, the driving laser wavelength is $\lambda=\dfrac{2\pi c}{\omega_{c}} =1064$nm and the mechanical quality
factor is $ Q $=6700. We also consider the resonant case $ \Delta=\omega_{m} $, i.e., when the optomechanical cooling generated by the laser. It should be pointed out that
depending on whether $ P/P_{c} <1 $ or $ P/P_{c} >1 $, where the critical pump power $ P_{c} $ is given by \cite{Huang6,Yan}
$ P_{c}=\dfrac{m\omega_{m}\omega_{c}(\kappa^{2}+\omega_{m}^{2})[\kappa-(\gamma_{m}/2)^{2}]}{4\kappa g_{0}^{2}} $, the optomechanical system works in the weak coupling
regime of optomechanically induced transparency (OIT) or in the strong coupling regime of normal-mode splitting(NMS).
For the above chosen experimental parameters, the critical power is about $ P_{c}=3.83 $ mW.
For $ P=5 $ mW and $ T=100 $ mK, we plot the displacement and momentum spectra of the
moving mirror versus the normalized frequency $ \omega/\omega_{m} $ in Fig. \ref{plot1}  for the cases when the finite-bandwidth squeezed vacuum input field is generated by DPO and NDPO, respectively. In each
figure, the result for the case of normal vacuum input $ (M=0, N=0) $ is also shown for comparison. As is seen from Fig.\ref{plot1}, in the case of squeezed vacuum from DPO, the spectrum $ S_{q}(\omega) $ has three peaks while $ S_{p}(\omega) $ has a single sharp peak centered at $ \omega=\omega_{m} $ . For the NDPO squeezed vacuum input noise, as shown in Fig. (\ref{plot1}(c), we can see a pimple and a hole in the displacement spectrum of the mirror which are induced by the two-photon correlation characteristic of the squeezed
vacuum input noise. The appearance of two peaks in the momentum spectrum [Fig. \ref{plot1}(d)] is a manifestation of the two symmetric peaks in the two-mode squeezed input noise that are separated by $  \alpha$ with respect to the carrier frequency $  \omega_{s}$.\\
\onecolumngrid
\vspace*{0mm}
\twocolumngrid

\begin{figure}[ht]
\includegraphics[width=150mm]{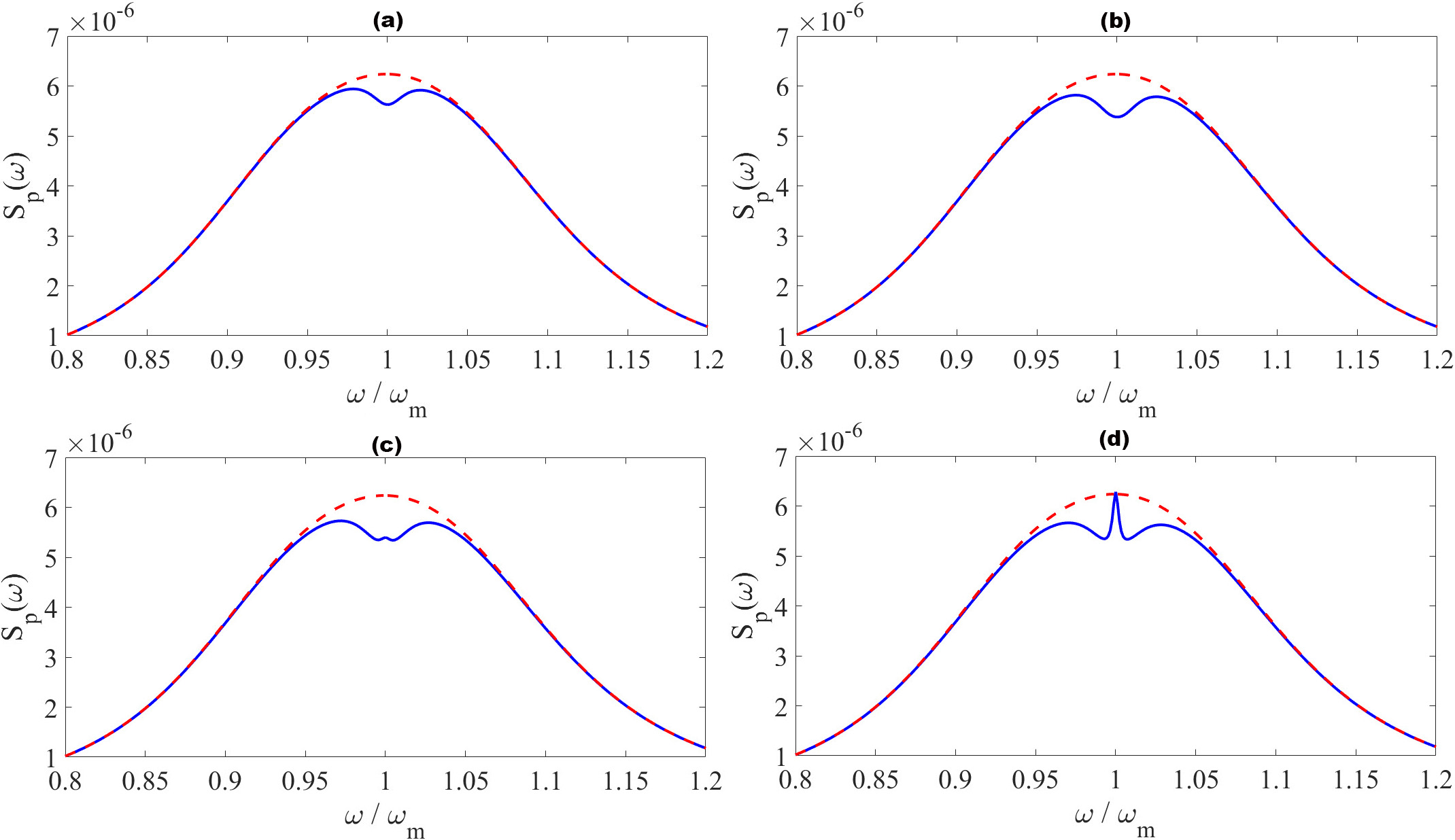}
\vspace*{-5mm}
\caption{\footnotesize (Color online) The momentum spectrum (W/Hz) of the moving mirror versus the normalized frequency  $ \omega / \omega_{m} $ for DPO (blue line)
with various values of the pumping rate $ \epsilon$: (a) $ 0.1\kappa_{p} $, (b) $ 0.2\kappa_{p} $, (c) $ 0.3\kappa_{p} $, (d)$ 0.4\kappa_{p} $. Here, we set $ \phi_{0}=\pi $. The spectrum(W/Hz) for the case of normal vacuum input is plotted for comparison (red dashed line). The other parameters are the
same as those in Fig. 2. }\label{plot3}
\end{figure}

\begin{figure}[ht]
\includegraphics[width=150mm]{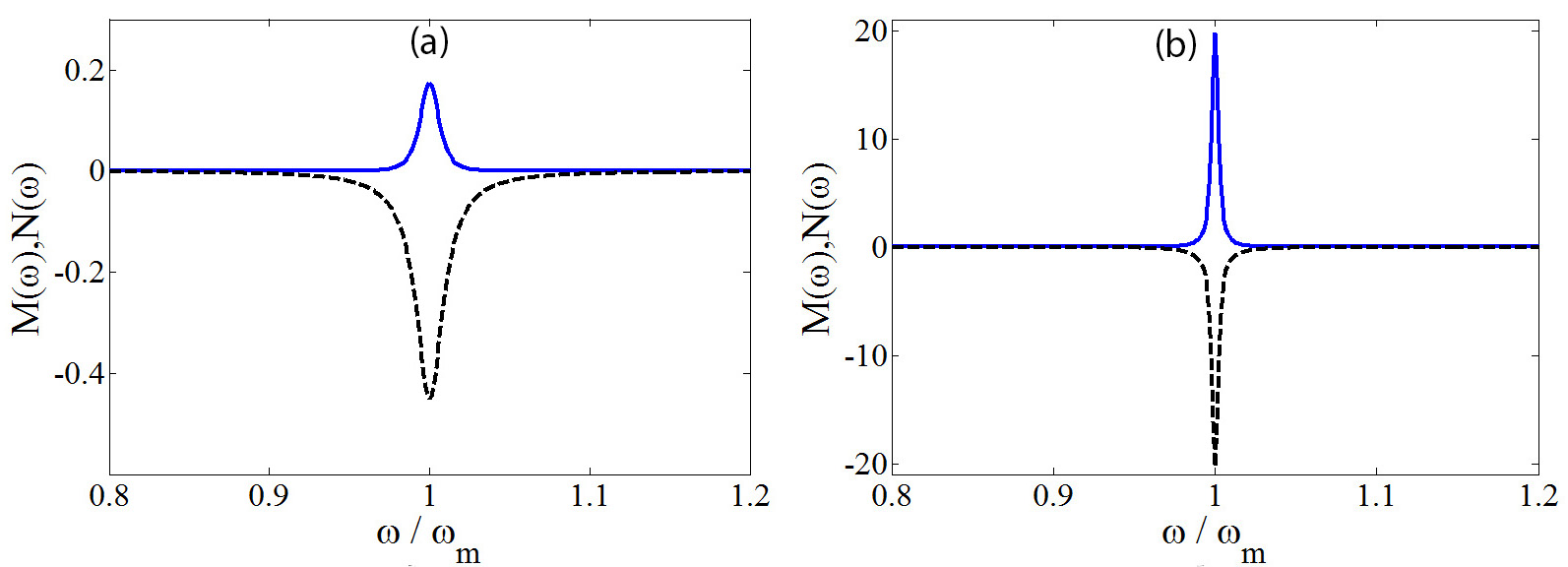}
\vspace*{-5mm}
\caption{\footnotesize (Color online) $ M(\omega) $ (dashed black line) and $ N(\omega)$ (blue line) versus the normalized frequency  $ \omega / \omega_{m} $ for DPO with the parameters of  (a) Fig. 3(a) and (b) Fig. 3(d). }\label{plot4}
\end{figure}

Another interesting feature appears for the same data as of Fig. (\ref{plot1}) but for $ \phi_{0}=\pi $. In Fig. (\ref{plot3}), for a fixed value of the DPO cavity decay rate ($ \kappa_{p}=0.1\kappa $), we change the pumping rate $ \epsilon$ which determines the intensity of the squeezed vacuum. As is evident from Figs. \ref{plot3}.(a) and \ref{plot3}.(b), the spectrum displays a visible dip at the line center; the larger the value of $ \epsilon$ (and so $  N(\omega)$), the deeper is the dip. The origin of the spectral hole burning is the negative contribution of the third term in  Eq. (\ref{final S_{p}}) which is proportional to $  M(\omega)$ and depends on $ \phi_{0}$. This feature may provide a way of detecting two-photon correlations in very weak fields. For more explanation, in Fig. \ref{plot4}(a) we have plotted  $  M(\omega)$ and $  N(\omega)$ for the parameters given in Fig. \ref{plot3}(a). The figure shows that the hole burning happens where $  M(\omega)$ is negative and has larger magnitude than $  N(\omega)$. This means that the correlation between two photons is more than the correlation of one photon with itself, which is due to the quantum nature of squeezed light. With increasing $ \epsilon$  slightly to $0.3\kappa_{p}  $ in Fig. \ref{plot3}(c), we see that the hole is replaced by a small pimple. The reason for this change in the spectral line is that as $  N(\omega)$ increases, its value gets
closer to the value of $  M(\omega)$ and the positive contribution of the second term of  Eq. (\ref{final S_{p}})  compensates the negative contribution (originated from the two-photon correlation characteristic of the squeezed vacuum input noise) or even becomes larger than it, as shown in  Fig. (\ref{plot3}.d). 
Although for $ \varepsilon=0.4\kappa_{p} $ Fig. \ref{plot4}(b) shows that $ |M(\omega)|=N(\omega) $ over the range of considered frequencies, the spectrum $ S_p(\omega) $ has a pimple [see Fig. \ref{plot3}(d)]. The reason for this is that due to the radiation pressure contribution, the term involving $ N $ [the second term in Eq. (\ref{final S_{p}})] becomes predominant over the term involving $ M $ [the third term in Eq. (\ref{final S_{p}})].
In the other case, if we consider $ \epsilon/\kappa_{p}=0.2 $, [similar to Fig. \ref{plot4}(b)], by increasing the bandwidth of input squeezed field (from $ 0.1\kappa $ to $ 0.3\kappa $),  the width of dip increases and finally the dip disappears.\\
\onecolumngrid
\vspace*{0mm}
\twocolumngrid

It is worth pointing out that the hole and pimple profiles of the spectrum are induced by the two-photon correlation
character of the squeezed vacuum. However, they are more visible if the thermal noise is reduced. One can show that the dip of the holes and the height of the pimples are modified with changing the environment temperature $ T $. As expected, if we increase the environment temperature, the unusual profiles disappear and, eventually, the spectral line becomes identical to the case associated with the normal vacuum injection.

\begin{figure}[ht]
\includegraphics[width=\textwidth]{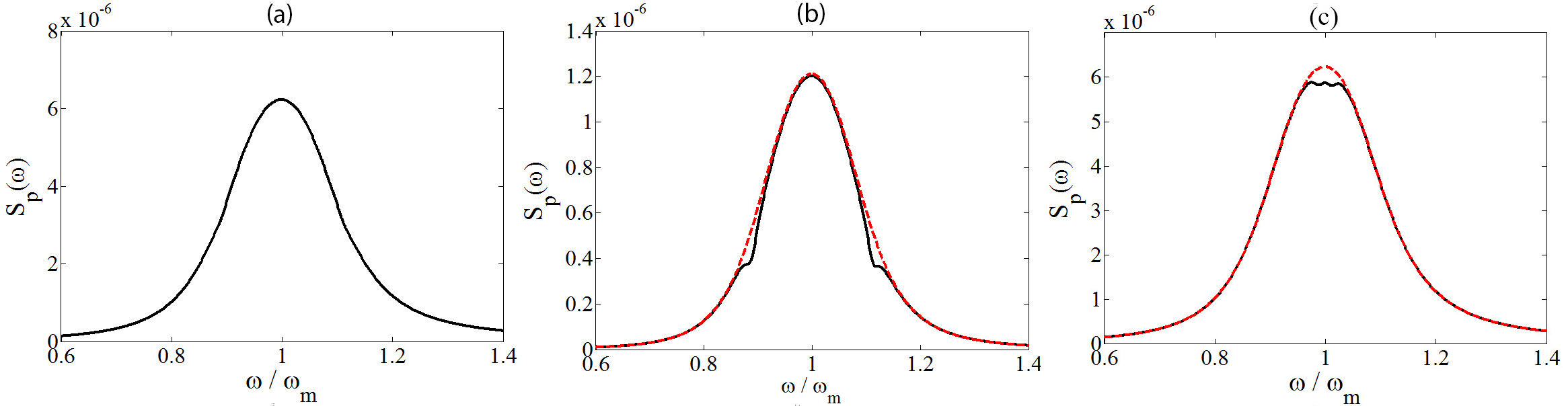}
\vspace*{-8mm}
\caption{\footnotesize (Color online)The momentum spectrum(W/Hz) of the moving mirror versus the normalized frequency $ \omega / \omega_{m} $ for the NDPO squeezed vacuum (black line) and the normal vacuum state (dashed red line) as input probe fields: (a) $ T=100$ mK, $\alpha=5\kappa_{p} $, (b) $ T=1$ mK, $\alpha=5\kappa_{p} $, (c) $ T=100$mK, $\alpha=0.5\kappa_{p} $. Other parameters are $\kappa _{p}=0.1\kappa,\epsilon=0.1\kappa_{p} ,\phi_{0}= \pi$, and $P=5$ mW.}\label{plot6}
\vspace*{1mm}
\end{figure}

In Fig. (\ref{plot6}), we can recognize  the manifestation of unusual shapes in the momentum spectral density for NDPO squeezed vacuum. 
As is seen from Fig. \ref{plot6}(a), at $ T=100 $ mK and for $ \alpha=5\kappa_{p} $, the two cases of normal vacuum and NDPO squeezed vacuum inputs result in a completely identical spectrum $ S_{p}(\omega) $. In this situation, the radiation pressure coupling is not so strong to overcome the effect of thermal noise of the
moving mirror. With decreasing the environment temperature to $ T=1 $ mK, the hole burning
happens and two holes appear around the central squeezing frequency $ \omega_{s} $. Although the
thermal noise effectively prevents the hole burning, with reducing the inter-mode frequency
separation, $ \alpha $, this phenomenon appears even when the temperature $ T  $ is raised [see Fig. \ref{plot6}(d)].

By the above numerical results, we have shown that how the quantum nature of finite-bandwidth squeezed
vacuum manifests itself in the unusual spectral features of the mirror. 

\begin{figure}[ht]
\includegraphics[width=\columnwidth]{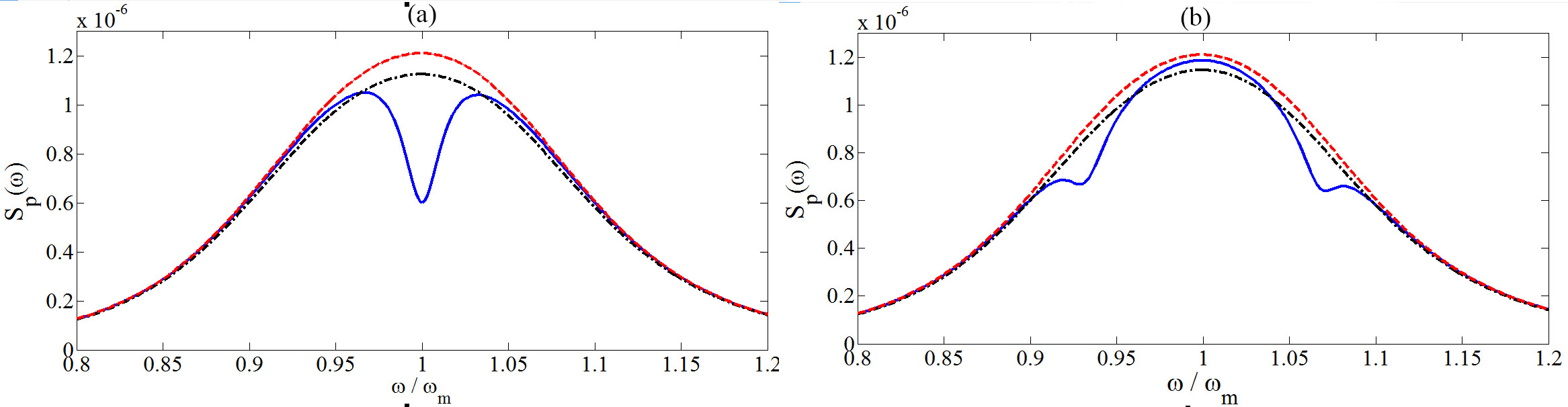}
\vspace*{-5mm}
\caption{\footnotesize (Color online) The momentum spectrum (W/Hz) of the moving mirror versus the normalized frequency  $ \omega / \omega_{m} $ for the case of finite  bandwidth squeezed vacuum with $ \kappa_{p}=0.1\kappa$ (blue line), infinite-bandwidth squeezed vacuum with $ \kappa_{p}=\kappa$ (black dashed- dotted line), and normal vacuum (dashed red line) for (a) DPO, and (b)NDPO . Here, we have set $\epsilon=0.01\kappa $, $\phi_{0}=\pi $,  $ \alpha=0.3\kappa $, $ T = 1\,{\rm mK} $, and $ P=5\,{\rm mW} $.Other parameters are the same as those in Fig.2. }\label{plot7}
\vspace*{1mm}
\end{figure}

At this point, it is interesting to address the question as to whether or not such features persist for the infinite-bandwidth case. For this purpose, we fix the maximum value of photon number $  N(\omega)$ (with $ \epsilon=0.01\kappa $) and compare the spectral features of the moving mirror for squeezed vacuum of finite-bandwidth ($ \kappa_{p}=0.1\kappa $) with those for infinite-bandwidth ($ \kappa_{p}=\kappa $) . In Fig. \ref{plot7}, the momentum spectrum of the moving mirror is plotted for both cases of DPO and NDPO squeezed vacuum inputs. As is seen, the spectral hole burning is not visible for the broad-band squeezed vacuum input.

\begin{figure}[ht]
\includegraphics[width=110mm]{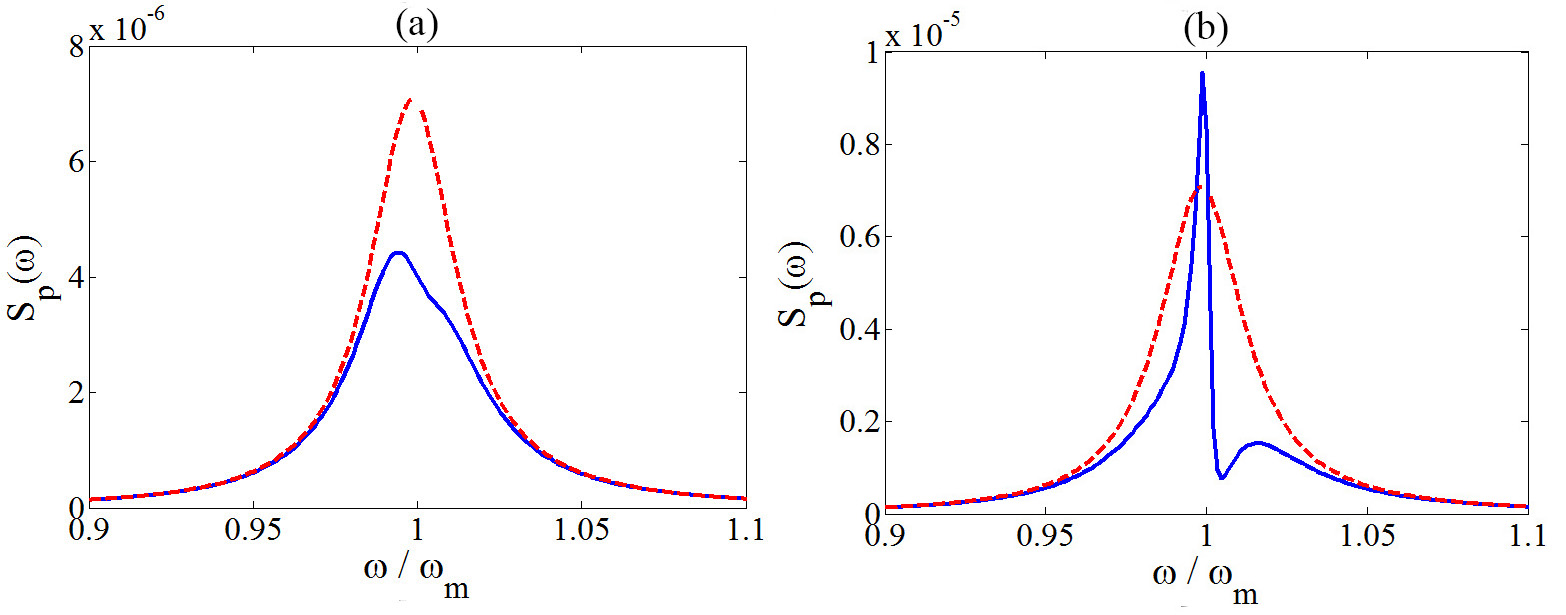}
\vspace*{-5mm}
\caption{\footnotesize (Color online) The momentum spectrum (W/Hz) of the moving mirror versus the normalized frequency $ \omega / \omega_{m} $ for the case of DPO squeezed vacuum with $ \phi_{0}=0.85\pi,\kappa_{p}=0.1\kappa $ and different values of $ |\epsilon| $: (a) $ 0.1\kappa_{p} $  (b) $ 0.4\kappa_{p} $ (blue line). The red dashed line shows the spectrum(W/Hz) for the case of normal squeezed vacuum input. Other parameters are the same as those in Fig.2.}\label{plot9}
\vspace*{-3mm}
\end{figure}

\begin{figure}[ht]
\includegraphics[width=180mm]{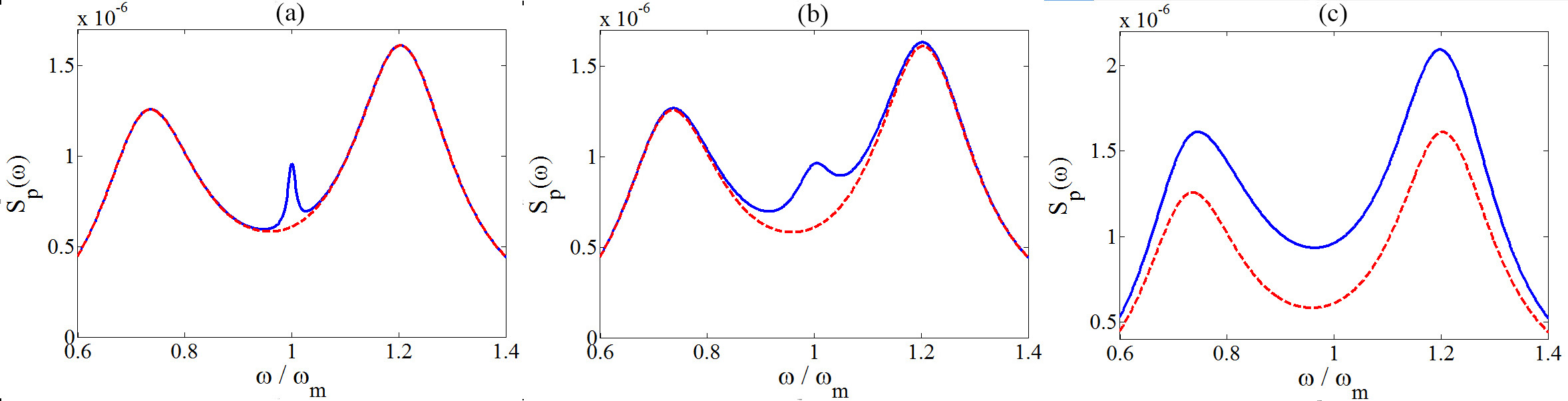}
\vspace*{-5mm}
\caption{\footnotesize (Color online) The momentum spectrum(W/Hz) of the moving mirror versus the normalized frequency $ \omega / \omega_{m} $ for DPO squeezed vacuum( blue line) as input probe field, with different values of $ \kappa_{p} $: (a) $0.1 \kappa $, (b) $ 0.5\kappa $, and (c) $ 10\kappa $ (broad-band squezeed vacuum). Other parameters are $ \epsilon=0.1\kappa _{p}, \phi_{0}=0, P=20\, {\rm mW},$ and $T=100 \, {\rm mK} $.The red dashed line shows the spectrum(W/Hz) for the case of normal squeezed vacuum input.}\label{plot10}
\vspace*{-1mm}
\end{figure}

\begin{figure}[H]
\includegraphics[width=170mm]{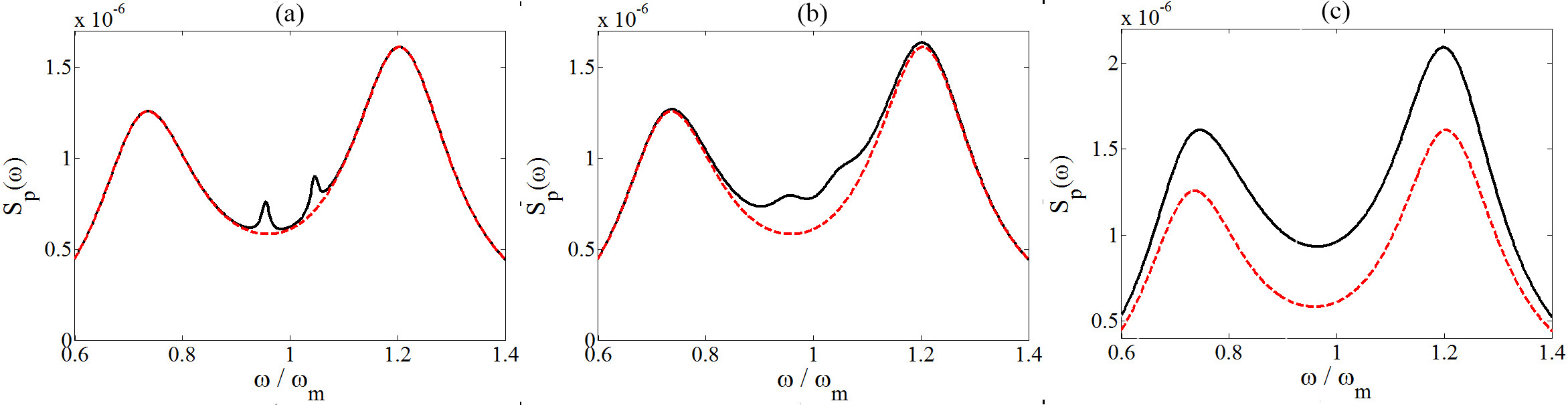}
\vspace*{-5mm}
\caption{\footnotesize (Color online) The momentum spectrum(W/Hz) of the moving mirror versus the normalized frequency $ \omega / \omega_{m} $ for  NDPO squeezed vacuum (black line) as input probe field for different values of $ \kappa_{p} $: (a) $0.1 \kappa $, (b) $ 0.5\kappa $, (c) $ 10\kappa $ (broad-band squeezed vacuum). Other parameters are $ \epsilon=0.1\kappa _{p},\alpha=0.2\kappa, \phi_{0}=0, P=20\,{\rm  mW},$ and $T=100\, {\rm mK} $.The red dashed line shows the spectrum(W/Hz) for the case of normal squeezed vacuum input.}\label{plot11}
\vspace*{-1mm}
\end{figure}

\onecolumngrid
\vspace*{0mm}
\begin{figure}[ht]
\includegraphics[width=\textwidth]{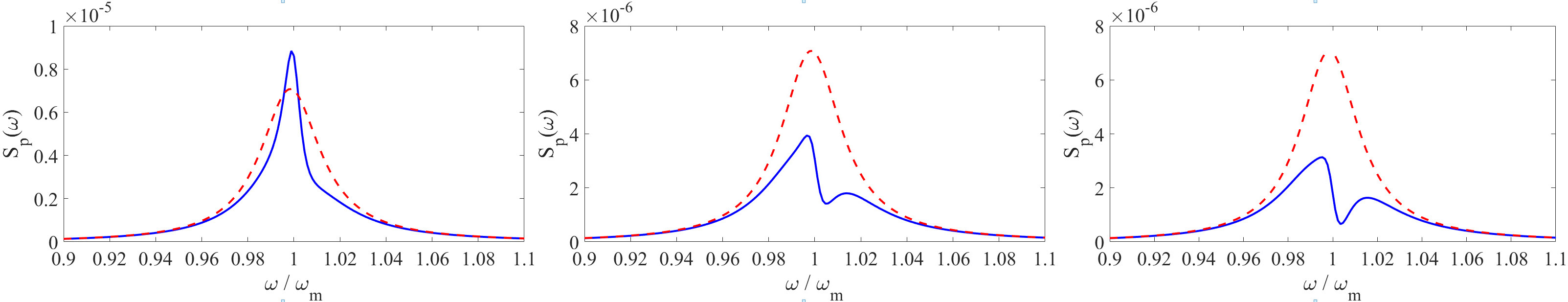}
\vspace*{-8mm}
\caption{\footnotesize (Color online) The momentum spectrum(W/Hz) of the moving mirror versus the normalized frequency 
$ \omega / \omega_{m} $ for the case of DPO squeezed vacuum input with
various values of the phase $  \phi_{0} $: (a) $ 0.75\pi $, (b) $ 0.85\pi $, and (c) $ 0.95\pi $ (blue line). The spectrum(W/Hz) for the case of normal vacuum input is plotted for comparison (red dashed line). Here, we have set $ \kappa_{p}=0.1\kappa $, $ \epsilon=0.3\kappa_{p} $, $T =1 $ mK and 
$ P = 1 $ mW. Other parameters are the same as those in Fig. 2.}\label{plot8}
\vspace*{0mm}
\end{figure}
\twocolumngrid
 In the investigation of interaction between a two-level atom and the squeezed vacuum, it has been shown\cite{Zhou1,Dalton,Swain2} that the fluorescence spectrum exhibits a dispersive-like profile which is associated with nonclassical characteristic of the squeezed vacuum. Here, we address the question as to whether or not such an anomalous feature can be observed in the spectra of the mechanical spectrum. The numerical analysis reveals that in the strong coupling regime of NMS neither the displacement nor momentum spectrum of the movable mirror exhibits the dispersive profile. However, in the weak coupling regime of OIT and when the thermal noise is small and quite negligible, the nonclassical nature of the squeezed vacuum can manifest itself in another anomalous spectral feature, i.e., the phase-sensitive narrow dispersive-like profile for the momentum spectrum of the mirror. This feature has been shown in  Fig. (\ref{plot8}) where we have plotted $ S_{p}(\omega) $ against $ \omega/\omega_{m} $ for the case of DPO squeezed vacuum input with different values of $ \phi_{0} $ and for $ P=1 $mW and $ T=1 $mK.
In Fig. (\ref{plot9}) we illustrate how the amplitude of the squeezed vacuum input affects the dispersive profile. In Fig. (\ref{plot9}.a) we do not see the unusual profile but with increasing$ |\epsilon| $, $ N $ (the photon number of squeezed vacuum) increases and dispersive profile appears. Numerical analysis shows that unlike the hole burning which is pronounced for small values of  $ N $, the dispersive profile appears for $ N>1 $. A similar result has been obtained for the interaction of a two-level atom with squeezed vacuum\cite{Dalton}. 
\onecolumngrid
\vspace*{-3mm}
\twocolumngrid
Now, we examine the response of the mechanical oscillator when the pump laser power $ P $ increases. Figures \ref{plot10} and \ref{plot11} show the effect of the damping rate of the parametric oscillator cavity, $ \kappa_{p} $ , on the momentum spectrum of the movable mirror, $ S_{p}(\omega) $, for DPO and NDPO squeezed vacuum inputs, respectively, when $ P=20 $ mW and $ \epsilon=0.1\kappa_{p} $. For the case of normal vacuum input (red dashed line) two-mode splitting is observed. This is because with increasing $ P $, the optomechanical coupling is strengthened and the usual NMS into two modes appears. Fig. \ref{plot10}(a) shows that when the optomechanical cavity is driven by a finite-bandwidth squeezed vacuum with small $ \kappa_{p} $ ($ \kappa_{p}=0.1\kappa $) a pimple appears at $ \omega=\omega_{m} $ in the momentum spectrum for the case of DPO squeezed vacuum input (NMS into three modes), while Fig.\ref{plot11}(a) indicates that for the NDPO case the spectrum exhibits two pimples around $ \omega=\omega_{m} $ (NMS into four modes). The three-mode (four-mode) splitting is associated with the mixing among the vibrational mode of the moving mirror, fluctuations of the cavity field around the steady state, and fluctuations of the single-mode DPO (two-mode NDPO) squeezed vacuum. With increasing $ \kappa_{p} $, the squeezed-vacuum bandwidth, and  the width of pimple increase and the pimples in both spectra associated with DPO and NDPO are suppressed [Figs. \ref{plot10}(b) and \ref{plot11}(b)]. In the broad-bandwidth limit ($ \kappa_{p}=10\kappa $), as is seen from Figs. \ref{plot10}(c) and \ref{plot11}(c), the pimples completely disappear and, as expected, the momentum spectra are identical for both cases of DPO and NDPO squeezed vacuum inputs.\\

\onecolumngrid
\vspace*{0mm}
\vspace*{0mm}
\twocolumngrid

\vspace*{-5mm}
\subsection{ Mechanical squeezing}
\begin{figure}[ht]
\includegraphics[width=130mm]{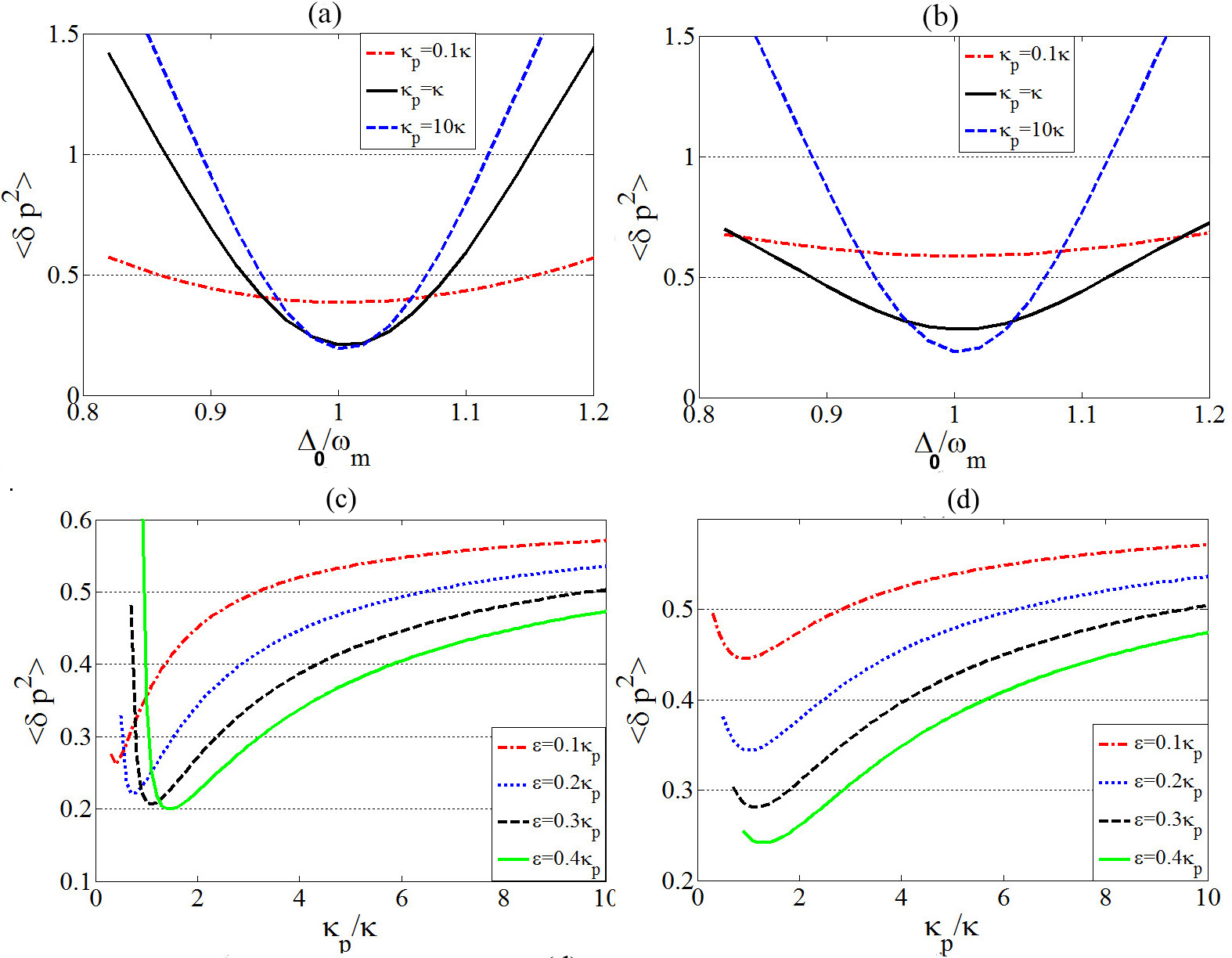}
\vspace*{-3mm}
\caption{\footnotesize The mean-square fluctuations in momentum of the movable mirror versus: the normalized detuning  $ \Delta_{0}/\omega_{m} $ for(a) DPO, and (b)NDPO squeezed vacuum as input probe fields with different values of $ \kappa_{p} $ (with $ \alpha=0.5\kappa,\epsilon=0.3\kappa_{p} $), and versus  $\dfrac{\kappa_{p}}{\kappa} $ for  (c) DPO (d) NDPO squeezed vacuum inputs with different value of $ \epsilon $. Here, we have set $\alpha=0.5\kappa, \phi_{0}=\pi, P=5 mW, T=1 mK $.} \label{plot16}
\vspace*{3mm}
\end{figure}

\begin{figure}[ht]
\includegraphics[width=\textwidth]{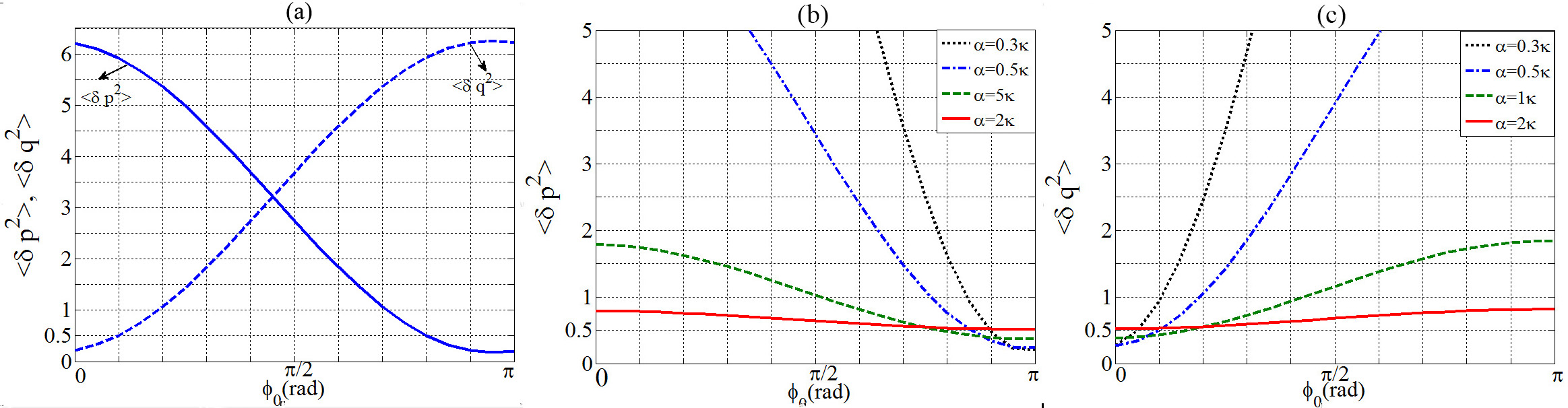}
\vspace*{-8mm}
\caption{\footnotesize (Color online) The mean-square fluctuations   in (a) displacement and momentum of the movable mirror for the case of DPO squeezed vacuum input with $\epsilon=0.3\kappa$, (b) momentum, and (c) displacement of the movable mirror for the case of NDPO squeezed vacuum input with $\epsilon=0.47\kappa,\alpha=0.5\kappa$, and different values of $ \alpha $ versus the squeezing phase $ \phi_{0} $. Other parameters are $ \kappa_{p}=\kappa,P=5$ mW, $T=1$ mK.}\label{plot18}
\vspace*{-1mm}
\end{figure}

\begin{figure}[ht]
\includegraphics[width=130mm]{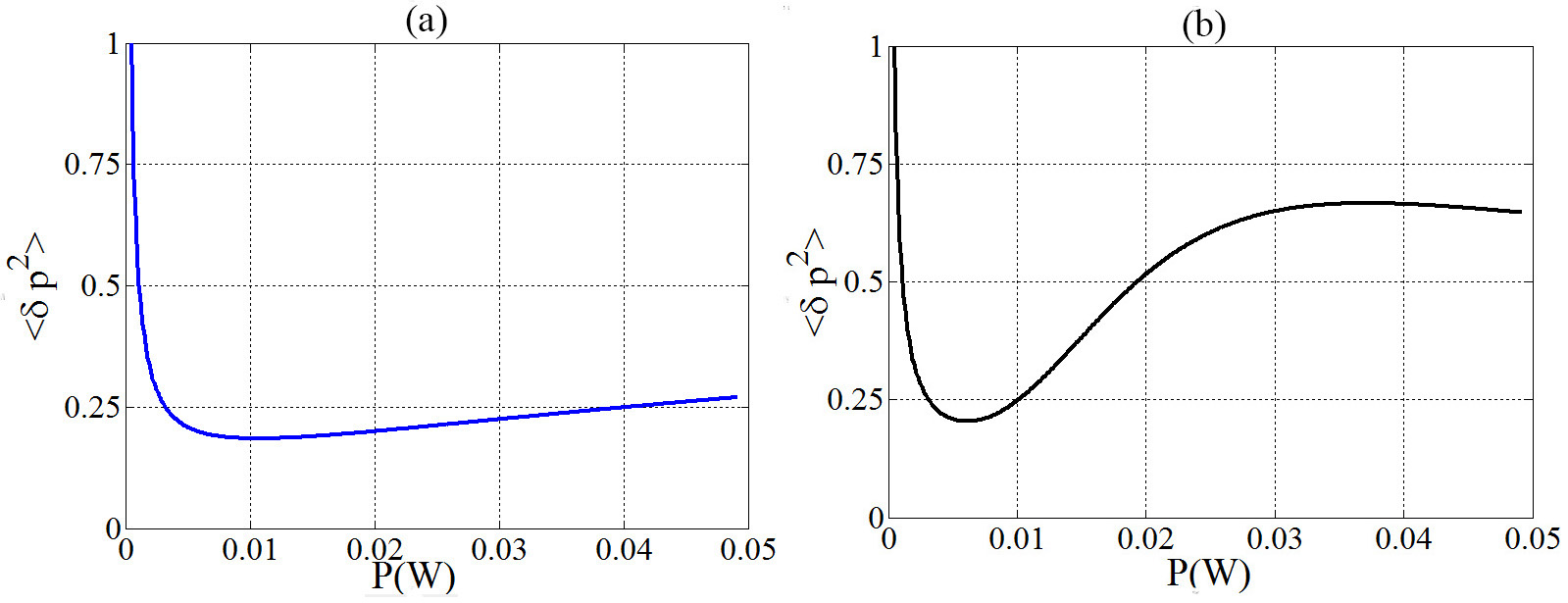}
\vspace*{-3mm}
\caption{\footnotesize (Color online) The mean-square fluctuations in momentum of the movable mirror versus pump laser power ($P $) for (a) DPO, and (b) NDPO squeezed vacuum input. Here, we have set $ \kappa_{p}=\kappa $, $ \epsilon=0.3\kappa, \alpha=0.3\kappa $, $ \phi_{0}=\pi $ and $ T=1$ mK.}\label{plot19}
\end{figure}

\vspace*{0mm}
In this subsection, we consider the quantum fluctuations in the momentum and displacement quadratures of the movable mirror to investigate the mechanical squeezing and its dependence on different parameters of the system under consideration. In Fig. \ref{plot16}, we have plotted the variance $\langle \delta p^2\rangle $ as a function of normalized detuning $ \Delta_{0}/\omega_{m} $ for $ \mu=0.2\kappa_{p} $ with different values of $\kappa_{p}$. As is seen from Fig. \ref{plot16}(a), in the case of DPO squeezed vacuum input the momentum squeezing occurs ($\langle \delta p^2\rangle<1/2  $) for both finite and infinite (similar to results in \cite{Huang2}) bandwidth cases. However, with increasing the squeezed vacuum bandwidth (i.e., increasing $\kappa_{p}$), the maximum amount of momentum squeezing increases, while the range of $ \Delta_{0} $ over which the momentum squeezing appears is decreased.  Furthermore, the figure shows that the optimal momentum squeezing is obtained via tuning $ \Delta_{0} $ around $ \omega_{m} $. Figure \ref{plot16}.b reveals that the same results hold for the case of NDPO squeezed vacuum input with the only difference that there is no momentum squeezing for finite-bandwidth squeezed vacuum excitation.  In Figs. \ref{plot16}(c) and \ref{plot16}(d), we have examined the effect of the amplitude $ \epsilon $ of the coherent field driving the parametric oscillator on the momentum squeezing of the movable mirror. Figures \ref{plot16}(c) and \ref{plot16}(d) illustrate the behavior of $\langle \delta p^2\rangle $ as a function of $\kappa_{p}/\kappa $ for  the DPO and NDPO squeezed vacuum inputs, respectively, with different values of $\epsilon$. As can be seen, for a given value of $\epsilon$, the optimal momentum squeezing occurs for $ \kappa_{p}<2\kappa $ and with increasing the ratio $\kappa_{p}/\kappa $ the momentum fluctuations increase (similar to results in \cite{Jahne}). Moreover, the momentum squeezing is enhanced as the squeezing level ($\epsilon/\kappa_{p}$) increases.\\   
      In Fig.(\ref{plot18}), the effect of the phase $ \phi_{0} $ of squeezed vacuum field on the mechanical squeezing of the movable mirror is illustrated. In Fig. \ref{plot18}(a), we present the plots of $\langle \delta p^2\rangle $ (solid line) and $ \langle \delta q^{2}\rangle  $ (dashed line) against  $ \phi_{0} $ when the optomechanical cavity is driven by DPO squeezed vacuum light. We find that the momentum squeezing appears for $ \phi_{0}>0.78\pi $ while the displacement quadrature exhibits squeezing for $\phi_{0}<0.12\pi $. Thus, controlling the phase $ \phi_{0} $  provides the possibility of squeezing transfer from the squeezed vacuum driving field to the momentum or displacement quadratures of the moving mirror. Figures \ref{plot18}(b) and \ref{plot18}(c) illustrate , respectively, $\langle \delta p^2\rangle $ and $ \langle \delta q^{2}\rangle  $ as functions of $ \phi_{0} $ for the case of NDPO squeezed vacuum input with different values of the parameter $\alpha$. As can be seen, for both momentum and displacement quadratures, the optimal squeezing appears for small values of $\alpha$; the smaller is the inter-mode separation, the less is the minimum value of mechanical fluctuations. In addition, similar to the case of DPO squeezed vacuum input, the momentum and displacement squeezing occur for small and large values of the phase $ \phi_{0} $, respectively. For the investigation of the squeezing level of mechanical oscillator, one can obtain the mechanical covariance matrix and find the variance of the maximally squeezed and anti-squeezed
quadratures as its smallest and largest eigenvalues \cite{Otey}.\\
  Finally, we examine the effect of the pump laser power, $ P $ , on the mechanical squeezing of the moving mirror. In Figs. \ref{plot19}(a) and \ref{plot19}(b), we have plotted $\langle \delta p^2\rangle $ versus the power $ P $ for the DPO and NDPO finite-bandwidth squeezed vacuum inputs, respectively. We find that with increasing the pump laser power the momentum fluctuations of the mirror decrease. The optimal momentum squeezing for the case of DPO (NDPO) squeezed vacuum is achieved around $ P=10$ mW ($ P=6$ mW). The figures also show that unlike the case of NDPO, the momentum squeezing persists at high powers of the pump laser when the optomechanical cavity is driven by DPO squeezed vacuum.

\onecolumngrid
\vspace*{1mm} 

\vspace*{0mm}

\vspace*{0mm}
\vspace*{10mm} 
\twocolumngrid
 
 \section{conclusions}\label{sec6}
In conclusion, we have investigated the response of a mechanical oscillator in an optomechanical cavity driven by a finite-bandwidth squeezed vacuum excitation generated by a DPO or a NDPO. By using the quantum noise approach, we have analyzed the effects of the bandwidth and squeezing parameters of the squeezed vacuum input on the displacement and momentum fluctuations spectra as well as the mechanical squeezing of the movable mirror. 
     Our results interestingly show that even for small squeezing bandwidths, the spectra of the mechanical oscillator exhibit anomalous features that are unique to the quantum nature of squeezed light. In this respect, we have shown that pimple, hole burning, narrowing of the spectral line and dispersive profile can be observed in the mechanical spectra. These phenomena have previously been observed in the spectra of a two-level atom interacting with squeezed vacuum but, to the best of our knowledge, it is the first time that such anomalous features are reported in cavity optomechanics. We have found that the squeezing parameters as well as the mirror temperature affect the hole burning, and the phase of driving squeezed vacuum plays a key role in the appearance of dispersive profile which appears at high intensities of the squeezed vacuum input. When the hole burning appears, the two-photon correlation of the driving squeezed vacuum is transferred to the mechanical spectra of the movable mirror. We have also found that in the case of finite-bandwidth NDPO squeezed vacuum input when the pump laser power increases two pimples appear in the momentum spectrum around the squeezing carrier frequency $ \omega_{s} $, indicating the two-mode nature of the driving squeezed vacuum. For the case of DPO squeezed vacuum input, only one pimple appears at frequency $ \omega_{s} $. These features strongly depend on the squeezing bandwidth and do not extend into the regime of broadband squeezed vacuum excitation.

 We have also studied the mean-square fluctuations in displacement and momentum of the movable mirror. In certain situations, the squeezing of the input probe field is transfered to the mirror. We have investigated the effect of intrinsic properties of squeezed light ($ \kappa_{p}, \epsilon, \alpha, \phi_{0} $) on the squeezing of the mechanical oscillator. It has been shown that the optimal squeezing occurs in the finite-bandwidth regime.
  In addition, the results show that depending on the value of the squeezing phase $\phi_{0} $, the input squeezing can be transferred into the momentum or displacement quadrature.  To sum up, the obtained results clearly show that an optomechanical system can be potentially be considered as a good candidate for detection and characterization of squeezed light.   
  
\end{document}